\documentclass[twocolumn]{aastex701}
\usepackage{CJKutf8}

\newcommand{\hetg}{$g_\mathrm{HETDEX}$}

\begin{document}

\title{Satellite Streaks in HETDEX: Spectra, Orbital Identification, and Survey Impact}


\author[0000-0002-2307-0146]{Erin {Mentuch Cooper}}
\affiliation{Department of Astronomy, The University of Texas at Austin, 2515 Speedway Boulevard, Austin, TX 78712, USA}
\affiliation{McDonald Observatory, The University of Texas at Austin, 2515 Speedway Boulevard, Austin, TX 78712, USA}
\email[show]{erin.hetdex@gmail.com}

\author[0000-0002-8433-8185]{Karl Gebhardt}
\affiliation{Department of Astronomy, The University of Texas at Austin, 2515 Speedway Boulevard, Austin, TX 78712, USA}
\email{gebhardt@astro.as.utexas.edu}

\author[0000-0002-8925-9769]{Dustin Davis}
\affiliation{Department of Astronomy, The University of Texas at Austin, 2515 Speedway Boulevard, Austin, TX 78712, USA}
\email{dustin@astro.as.utexas.edu}

\author[0000-0003-2575-0652]{Daniel J. Farrow}
\affiliation{E. A. Milne Centre for Astrophysics
University of Hull, Cottingham Road, Hull, HU6 7RX, UK}
\affiliation{Centre of Excellence for Data Science,
Artificial Intelligence \& Modelling (DAIM),
University of Hull, Cottingham Road, Hull, HU6 7RX, UK}
\email{D.J.Farrow@hull.ac.uk}

\author[0000-0001-7010-7637]{Barbara G. Castanheira}
\affiliation{Department of Physics and Astronomy, Baylor University, Waco, TX 76798-7316}
\email{Barbara_Endl@baylor.edu}

\author[0000-0002-3031-2588]{Gavin Dalton}
\affiliation{Department of Physics, Denys Wilkinson Building, Keble Road, Oxford, OX1 3RH, UK}
\email{gavin.dalton@physics.ox.ac.uk}

\author[0000-0001-6717-7685]{Gary J. Hill}
\affiliation{McDonald Observatory, The University of Texas at Austin, 2515 Speedway Boulevard, Austin, TX 78712, USA}
\affiliation{Department of Astronomy, The University of Texas at Austin, 2515 Speedway Boulevard, Austin, TX 78712, USA}
\email{hill@astro.as.utexas.edu}

\author[0009-0007-1326-6712]{Stephen C. Hummel}
\affiliation{McDonald Observatory, The University of Texas at Austin, 2515 Speedway Boulevard, Austin, TX 78712, USA}
\email{stephenhummel@utexas.edu}

\author[0000-0001-5561-2010]{Chenxu Liu\begin{CJK*}{UTF8}{gkai} (刘辰旭) \end{CJK*}}
\affiliation{South-Western Institute for Astronomy Research, Key Laboratory of Survey Science of Yunnan Province, Yunnan University, Kunming, Yunnan 650500, People's Republic of China}
\email{lorenaustc@gmail.com}

\author[0000-0003-3823-8279]{Shiro Mukae}
\affiliation{Department of Physics, School of Advanced Science and Engineering, Faculty of Science and Engineering, Waseda University, 3-4-1 Okubo, Shinjuku, Tokyo 169-8555, Japan}
\affiliation{Department of Astronomy, The University of Texas at Austin, 2515 Speedway Boulevard, Austin, TX 78712, USA}
\email{shiromukae88@gmail.com}

\author[0000-0002-6186-5476]{Shun Saito}
\affiliation{Institute for Multi-messenger Astrophysics and Cosmology, Department of Physics, Missouri University of Science and Technology, 1315 N. Pine St., Rolla MO 65409, USA}
\affiliation{Kavli Institute for the Physics and Mathematics of the Universe (WPI), The University of Tokyo Institutes for Advanced Study (UTIAS), The University of Tokyo, Chiba 277-8583, Japan}
\email{saitos@mst.edu}

\author[0000-0001-7240-7449]{Donald P. Schneider}
\affiliation{Department of Astronomy \& Astrophysics, The Pennsylvania State University, University Park, PA 16802, USA}
\affiliation{Institute for Gravitation and the Cosmos, The Pennsylvania State University, University Park, PA 16802, USA}
\email{dps7@psu.edu}

\author[0000-0002-7327-565X]{Sarah Tuttle}
\affiliation{Department of Astronomy, University of Washington, Seattle, 3910 15th Ave NE, Room C319, Seattle WA 98195-0002}
\email{tuttlese@uw.edu}

\begin{abstract}

We present a spectroscopic database and analysis of satellite streaks in Public Data Release 1 (PDR1) from the Hobby-Eberly Telescope Dark Energy Experiment (HETDEX), covering 6778 observations and $\sim$90\,deg$^2$ of sky from January 2017 to July 2024. We detect 527 streaks across 492 observations, a contamination fraction of 7.3\% that rises to monthly values above 20\% by 2023--2024, although the streaks mask only 0.09\% of the survey area. Of the 527 streaks, 475 (90.1\%) are identified with a cataloged object using archival two-line elements: 258 (54\%) lie in low or medium Earth orbit and 217 (46\%) are geostationary or highly elliptical. Only 35\% were launched since 2010; the rest are spent rocket bodies, debris, and older spacecraft. This reflects HETDEX's sensitivity to faint
objects and its 2017--2024 observing baseline. The Hobby-Eberly Telescope's large 10\,m aperture and HETDEX's $\sim$6\,minute exposures result in similar observed magnitudes for low Earth orbit (LEO) and geostationary Earth orbit (GEO) objects (median $g_{\rm AB}=16.2\pm1.1$ for LEO, $16.7\pm1.1$ for GEO). However, once transit time is corrected for, instantaneous magnitudes differ widely: $g_{\rm inst}=7.4$ for LEO versus $g_{\rm inst}=13.0$ for GEO objects, a difference of 5.6\,mag, while the exposure-averaged magnitudes differ by only 0.5\,mag. Normalized to 550\,km, the 34 Starlink streaks give $g(550\,{\rm km})=5.95$, consistent with the mean $V(550\,{\rm km})=5.92\pm0.04$ reported for VisorSat-design spacecraft. The spectra are consistent with reflected sunlight, with modest, signal-to-noise-driven differences in slope. The catalog, spectra, and code are archived at
\dataset[10.5281/zenodo.22962479]{https://doi.org/10.5281/zenodo.22962479}.

\end{abstract}


\section{Introduction} 

Over the past decade, the rapid deployment of large satellite constellations has introduced a growing and now unavoidable source of contamination in astronomical observations. SpaceX's Starlink constellation now exceeds 10,000 active satellites in low Earth orbit (LEO), comprising roughly two-thirds of all satellites currently operating, with tens of thousands more planned from competing constellations worldwide\footnote{\url{https://www.scientificamerican.com/article/spacex-reaches-milestone-of-10-000-starlink-satellites-in-orbit/}}. Because satellites reflect sunlight, they produce bright, moving trails in astronomical images that can render exposures unusable or introduce scattered light and increased background noise. 

Early concerns about satellite streaks affecting wide-field imaging surveys have since been confirmed by both simulations (e.g., \citealt{McDowell2020,Tyson2020}) and observational studies (e.g., \citealt{Hainaut2020}; \citealt{Rawls2020}; \citealt{Hainaut2026}). For next-generation wide-field facilities such as the Vera C. Rubin Observatory, these studies predict that even for constellations smaller than those currently proposed, satellite trails will affect $\sim$20\% of exposures at midnight and as many as 30–80\% during twilight (\citealt{Tyson2020}; \citealt{Hainaut2026}).

This rapid population and pollution of LEO poses a growing threat to astronomy. Recent simulations forecast that completion of proposed mega constellations could contaminate over one-third of Hubble Space Telescope exposures and more than 96\% of images from next-generation space telescopes \citep{Borlaff2025}. Recent observations from the SPHEREx space telescope have confirmed these predictions, with satellite trails detected in 73.3\% of exposures and an average of 2.18 trails per exposure \citep{borlaff2026}.

Most existing studies have focused on imaging surveys, where satellite trails primarily affect a fraction of pixels that can, in principle, be masked or interpolated. A recent census of 32 million wide-field
photometric observations from the Mini-MegaTORTORA network quantifies this
directly, finding that satellites, rockets, and debris brighter than IAU
interference thresholds \citep{IAU2024} already occupy a measurable fraction of the sky,
and forecasting that the roughly hundredfold growth projected for the
LEO constellation population would push contamination
probabilities for short, wide-field exposures toward unity
\citep{Mallama2026}. Empirical constraints on the brightness, frequency, and spectral properties of satellite streaks remain limited. In integral-field spectroscopic data, satellite streaks introduce spatially extended, wavelength-dependent contamination that cannot be trivially removed and may propagate into source catalogs, biasing both detections and derived physical properties. This is particularly relevant for wide-field spectroscopic surveys such as the Hobby–Eberly Telescope Dark Energy Experiment (HETDEX, \citealt{Gebhardt2021, Hill2021}), which uses untargeted emission-line searches over large sky areas to construct massive samples of galaxies at $1.9 < z < 3.5$.

In this paper, we present a spectral database and analysis of satellite streaks in integral-field unit (IFU) data from the Public Data Release 1 of HETDEX \citep[PDR1;][]{hetdex_pdr1}. In Section\,\ref{sec:data}, we provide a brief overview of PDR1 and describe the satellite detection, spectral extraction, and orbital crossmatching methods used to build the catalog. In Section\,\ref{sec:results}, we present the spectral database, including measurements of integrated brightnesses, surface brightnesses, and spectra over the wavelength range 3500–5500\,\AA. We summarize orbital catalog match properties and show that satellite contamination is already a common feature in HETDEX observations and that its incidence scales with the increasing number of satellites in orbit. All positions reported in this paper are in the International Celestial Reference System (ICRS; epoch J2000)\null.  All magnitudes are expressed in the AB system \citep{oke1983}. Spectral data are provided in air in the wavelength range from 3470\,\AA\ to 5540\,\AA\ in steps of 2\,\AA. 

\section{Data and Satellite Identification}
\label{sec:data}

\begin{figure*}[t]
    \centering
    \includegraphics[width=0.95\textwidth]{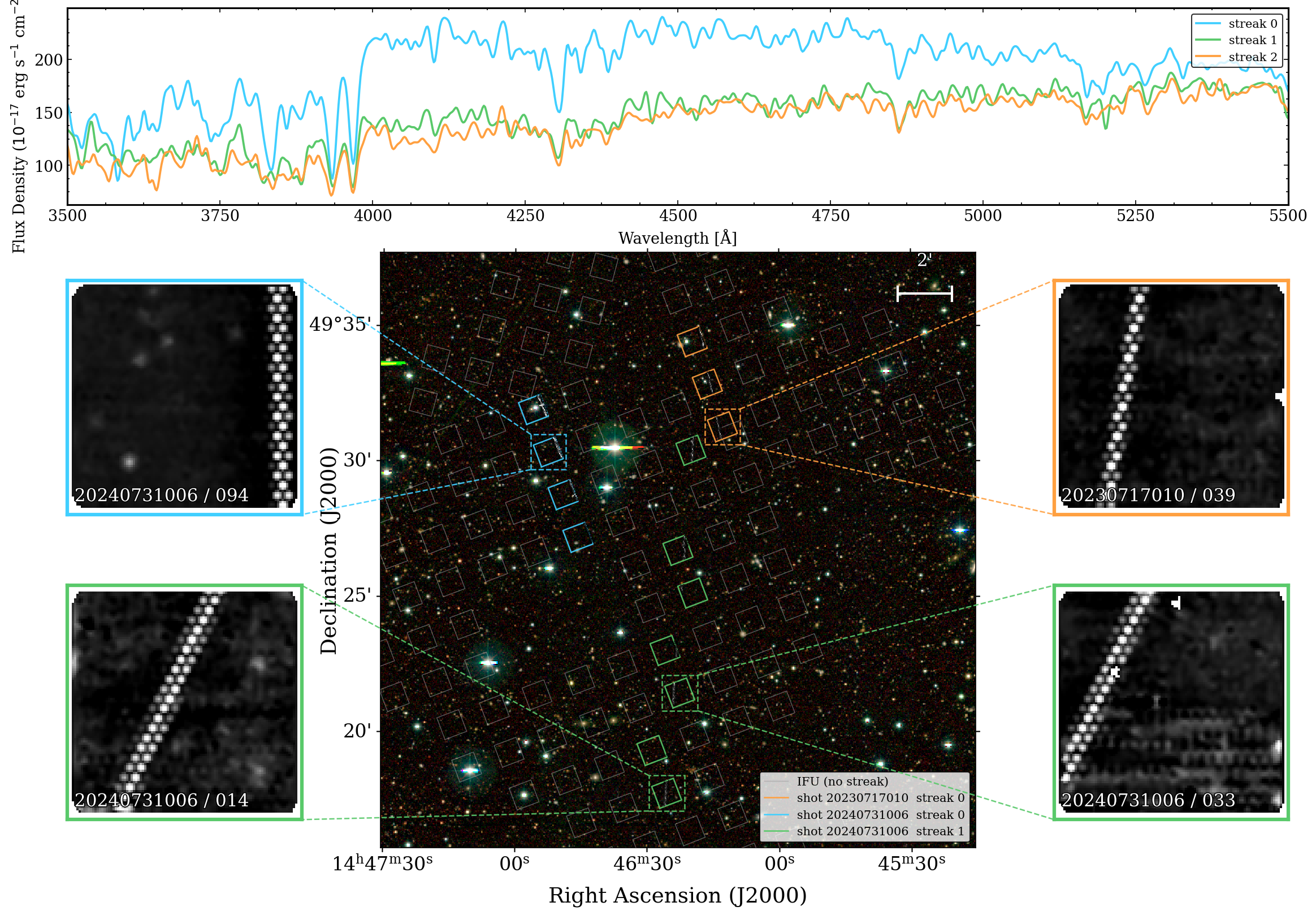}
    \caption{Satellite-streak detections in HETDEX overlaid on a Legacy Survey DR9 $grz$ image ($\approx$22\arcmin$\times$22\arcmin; RA$\,=\,221.596^{\circ}$,
    Dec$\,=\,+49.445^{\circ}$). IFU footprints are colour-coded by satellite trail: orange (shotid=20230717010), cyan and green (two separate streaks in shotid=20240731006). Zoom panels show four IFUs (20240731006/094, 20240731006/014, 20230717010/039, 20240731006/033) with their spaxels spectrally collapsed onto an image, revealing the characteristic dotted-line pattern of satellite trails sampled by the three-dither exposure fiber grid. Top: summed spectra over all IFUs in an observation for each streak. The spectra are consistent with reflected sunlight from LEO satellites.}
    \label{fig:satellite_streaks}
\end{figure*}

HETDEX is a wide-field integral-field spectroscopic survey designed to map the large-scale structure of the Universe at $z\sim2.4$ \citep{Gebhardt2021}. HETDEX obtains low-resolving-power ($R\sim800$--950) optical spectroscopy over 3500--5500\,\AA\ using the Visible Integral-field Replicable Unit Spectrograph (VIRUS, \citealt{Hill2021}) instrument on the Hobby-Eberly Telescope (HET, \citealt{Ramsey1998, Hill2021}), which consists of an array of replicated IFUs. Each HETDEX observation is composed of three-dithered exposures that together provide nearly complete spatial coverage within an individual $51\arcsec \times 51\arcsec$ IFU. They are referenced by their ``shotid'' in the format of \texttt{YYYYMMDDXXX} where \texttt{XXX} is the daily running index, ``obsid'', of observations. The layout of the IFUs in an observation contains gaps, such that only 1/4.7 of the HET's $\sim$22\arcmin\ field of view has IFU coverage in a single three-dither HETDEX observation. An example of the non-contiguous layout of IFUs in the focal plane is demonstrated in Figure\,\ref{fig:satellite_streaks} where HETDEX IFU regions are overplotted in light grey on a three-colour Legacy Survey DR9 $grz$ image \citep{Dey2019}. 

The fundamental internal data products are flux-calibrated fiber spectra, each 1\farcs5 in diameter, and a detections database generated directly from these data. Object detection is performed without imaging pre-selection using complementary searches for emission-line and continuum sources across more than $6\times10^8$ spectra. A full description of the detection methods can be found in \citet{Gebhardt2021}. For this study, only continuum sources are used for identification.  

The HETDEX continuum detection search assumes an object is point-source like. A detection is marked as a continuum source by measuring fiber counts in the PSF-extracted spectra in two 200\,\AA\ wide spectral regions, one in the blue (3700–3900\,\AA) and one in the red (5100–5300\,\AA). If either region exceeds 50 counts per 2\,\AA\ pixel, the fiber is stored as a continuum detection. This translates to roughly \hetg~$\sim22.5$, where \hetg\ is a pseudo-magnitude derived from the HETDEX source spectrum by convolving it with the SDSS $g$-band transmission curve and integrating the resulting flux.

\subsection{HETDEX Public Data Release 1}
\label{sec:pdr1}

This analysis makes use of the HETDEX Public Data Release 1 (PDR1; \citealt{hetdex_pdr1}), which provides both calibrated data cubes and associated source catalogs for the full HETDEX survey. The data cubes are constructed on a regular grid with 0\farcs5 spatial sampling and 2\,\AA\ spectral bins, covering 3500--5500\,\AA. Each cube spans $52\arcsec\times52\arcsec$ spatially and is interpolated from all IFU fibers within the three-dithered exposures. The data in the cubes consist of interpolated flux spatial pixels (spaxels), error spaxels and an associated bitmask for each spaxel in the data cube. A full description of the creation of the data cubes and the associated bitmasking is found in \citet{hetdex_pdr1}. 

The PDR1 catalogs are built from the internal HETDEX detection database, which includes both emission-line and continuum detections. These detections are grouped into astrophysical sources and assigned classifications and redshifts as described in \citet{Cooper2023} using a combination of spectral fitting and machine-learning techniques \citep{Davis2023, Debski2025, mukae2026}. In addition to astrophysical sources, the detection database contains a significant number of artifacts, including those arising from satellite streaks, which are identified and masked as part of the data reduction pipeline.

\subsection{Satellite Detection}

Satellite streaks are detected from spectrally collapsed image cutouts classified using the proprietary AI-classification software Rapid Automatic Image Categorization (RAIC; \citealt{raiclabs}). Continuum detections are converted into spectrally collapsed image cutouts by spectrally summing the PDR1 IFU data cubes from 4000 to 5000 \AA\null. The generated images are $20\arcsec\times20\arcsec$ and are centered on each continuum detection. For an example of these input images, please see examples in the right side of Figure~11 in \citet{hetdex_pdr1}.

Starting with an initial random sample set of $\sim$20K spectrally collapsed images, the images are uploaded into the RAIC classification framework. Labels are assigned to images with similar morphologies until each set has at least 1000 labeled samples. These are then applied as supervised inputs to the RAIC system for training. This trained model then assigns a probability for each label, including ``satellite'' and those exceeding a threshold classification score of 0.3 are flagged accordingly in our catalogs \citep{hetdex_pdr1}.
 
From these labels, satellite streaks are identified by searching observations that contain at least five continuum detections assigned a ``satellite'' label by RAIC. A satellite streak is confirmed when a group of detections in an observation displays an alignment of point-like detections tracing a linear path across the field. For these aligned satellite detections, the fitting is done manually by performing a linear fit in sky positions. Following the HETDEX masking strategy, detections and fibers within a fixed perpendicular angular distance of 6\,arcsec of the fitted track are associated with the satellite streak \citep{hetdex_pdr1}.  

An example of satellite streak detections in HETDEX is shown in Figure~\ref{fig:satellite_streaks}. When projected onto imaging data, satellite contamination appears as linear features spanning multiple IFUs across the $\sim$22\arcmin\ field of view. Due to the fiber-based, three-dither observing strategy of HETDEX, these streaks, which only strike through a single dither, produce a characteristic honeycomb pattern in the interpolated images. As in Figure\,\ref{fig:satellite_streaks}, multiple independent streaks can be present within a single observation, each corresponding to a distinct satellite track. These fits were performed manually by visually selecting tracks and then fitting each streak independently. The result for each linear fit is stored in the \texttt{hetdex\_api} Github repository \footnote{ \url{https://github.com/HETDEX/hetdex_api/blob/master/known_issues/hdr3/satellite_tracks.txt}}. We note that five observations within this list (20190731019, 20200523028, 20220928016, 20230720009, and 20240311025) are manually removed from the analysis because the spectra revealed the emission was due to a meteor. 

\subsection{Spectral Extraction and Photometric Measurements}
\label{sec:extractions}

For each identified satellite streak, we extract a one-dimensional spectrum by summing the flux of all spaxels whose sky positions fall within a fixed perpendicular distance of the fitted streak track. Following the HETDEX masking convention, this half-width is set to $6\arcsec$ (the \texttt{streaksize} parameter), so that spaxels lying within $\pm6\arcsec$ of the linear track---a strip $12\arcsec$ wide---are associated with the streak. Summing over this strip combines flux consistently across the multiple IFUs that a single streak crosses and avoids fragmenting the signal. The same set of associated spaxels defines the on-sky area used for the surface-brightness and masked-area measurements described below; because the strip width is fixed by the adopted half-width, the extracted area (and hence the surface brightness) scales with this choice.

Each spaxel in a HETDEX data cube is interpolated from the fibers of all three
dithered exposures (Section\,\ref{sec:pdr1}), so its flux is effectively an
average over the dither set. An astrophysical source that is present in every dither,
is recovered correctly; but a satellite crosses the field for only a single dither. Thus, we recover it by scaling
the summed flux and its propagated uncertainty by $N_{\rm exp}=3$ before
synthesizing photometry, which brightens the streak by
$2.5\log_{10}3 = 1.19$\,mag. 

Synthetic photometry is computed using the SDSS $g$-band filter as implemented in the \texttt{speclite} package \citep{speclite_pkg}. Each extracted spectrum is zero-padded to span the full range of the filter transmission curve, and an AB magnitude is computed from the flux-calibrated spectrum. All fluxes are expressed in units of $10^{-17}$\,erg\,s$^{-1}$\,cm$^{-2}$\,\AA$^{-1}$. The surface brightness of each streak is computed as the AB magnitude of the integrated flux divided by the projected on-sky area of the associated spaxels, yielding values in mag\,arcsec$^{-2}$.

\subsection{Orbital Crossmatching}
\label{sec:crossmatch}

For each streak we search for a catalogued object whose propagated orbit
reproduces the observed track, using archival two-line elements (TLEs) from the
Space-Track \texttt{gp\_history} service,\footnote{
\url{https://www.space-track.org}} which is the only source spanning the full
2017--2024 PDR1 baseline, since CelesTrak's public feed serves current elements
only and the IAU CPS SatChecker archive \citep{satchecker} begins in mid-2019.
Elements are propagated with the \texttt{sgp4} implementation of the SGP4/SDP4
theory \citep{sgp4pkg, vallado} over the intervals in which the shutter was
open. 

We note that we do not take these times directly from PDR1. The shot start times (column \texttt{mjd}) distributed
with the public release are stored at single precision, which near
${\rm MJD}=60{,}000$ quantises them to $2^{-8}$\,d, or 337.5\,s. This is comparable
to a single dither, and if used has the potential to miss crossings at the beginning of the observation. Instead per-dither shutter open and close times are read from the
internal HETDEX survey file (from internal data release HDR5, from which PDR1 is drawn), which covers all 492 shots in this sample,
and each of the three dithers is searched over
$[\mathrm{MJD}_{\rm open}-60\,{\rm s},\ \mathrm{MJD}_{\rm close}+60\,{\rm s}]$,
which comfortably exceeds the $\sim$35--40\,s of readout and repositioning
between dithers. The elapsed shot span ranges from 1169 to 2263\,s. These times are tabulated as \texttt{MJDopen} and \texttt{MJDclos},
and the \texttt{MJD} column carries the shutter-open time of the first
dither (Table~\ref{tab:catalog_columns}).

A coarse pass propagates every cataloged object with an element set for that night on a 20\,s grid with a median of $\sim$21{,}400 objects per shot, rising
to $\sim$25{,}000 by 2024 as the public catalog grows. This retains any object approaching within a search radius widened per object by half its own intersample motion; without this, a fixed radius undersamples fast LEO
objects, which move at up to $\sim$0.8\,deg\,s$^{-1}$. Survivors are
repropagated at 0.5\,s. SGP4 positions are rotated once per shot from TEME to GCRS --- whose axes match ICRS at the relevant precision --- and the observer's geocentric position subtracted to give topocentric directions directly comparable to the streak astrometry. No annual-aberration correction is applied. The HETDEX astrometric solution is tied to catalog star positions, so the same aberration is present in both frames, leaving a residual below 0\farcs1.

Since TLE error is dominated by its along-track component, with cross-track error at LEO of a few arcseconds to $\sim$1\arcmin, the primary discriminant is the perpendicular offset of the streak midpoint from the great circle locally tangent to the propagated track, together with the position-angle difference. We further require closest approach within $\pm2$\,s in time, obtained by decomposing the total separation into cross- and along-track components and converting the latter to a time offset via the local angular rate; this rejects objects lying along the same great circle but elsewhere on it. Candidates are accepted below $900\arcsec$ offset and $6\arcdeg$ in position angle, ranked by $(\mathrm{offset}/120\arcsec)^2 + (\Delta{\rm PA}/1\fdg5)^2$, and the best plus four runners-up are retained. A match is unambiguous when no runner-up exists or the next-best score exceeds three times the best plus one.

As a purity check, we evaluate each matched object's illumination at crossing with a conical Earth-shadow model \citep{vallado}: A match placing the object in Earth's umbra is physically impossible and sets a lower limit on the false-positive rate (Section~\ref{sec:identification}). Orbit classes follow from the propagated perigee $h_p$, apogee $h_a$ and eccentricity $e$: LEO ($h_a<2000$\,km), geostationary Earth orbit (GEO; $h_p>34{,}000$\,km and $h_a<37{,}500$\,km), highly elliptical orbit (HEO; $e>0.25$ with $h_a>2000$\,km, or $h_a\geq37{,}500$\,km), and medium Earth orbit (MEO) otherwise. Streaks with no acceptable candidate are additionally queried against SatChecker, which draws on a separate element archive; identifications recovered this way are flagged by \texttt{IDsrc} (Table~\ref{tab:catalog_columns}).

The per-dither shutter times supply a second, independent check. Each streak
carries the dither in which it was visually identified, taken from the manual
track fits, while its crossing epoch is set by SGP4 propagation of the matched
object; the two are derived independently. For 391 of the 469 matched streaks
with an assigned dither (83.4\%) the propagated crossing epoch falls inside
that dither's shutter interval, at a median of 60\% through the exposure. The
remainder are consistent with TLE along-track error and with streaks spanning
more than one dither, and show no systematic offset.

\subsection{Instantaneous Magnitude}
\label{sec:photometry}

The magnitude $g_{\rm AB}$ of Section~\ref{sec:extractions} is not the
satellite's instantaneous brightness: It is synthesized from a spectrum
calibrated as though the source shone for the full 360\,s HETDEX exposure
(after the dither-dilution correction of Section~\ref{sec:extractions}, which
adjusts the flux by a factor of three), whereas the satellite illuminates the summed spaxels only while crossing them.

For streaks with an orbital crossmatched counterpart, the topocentric angular rate
$\dot\theta$ at closest approach sets the rate at which the object crosses the VIRUS focal plane. Only the fraction of the streak footprint actually covered by IFU
spaxels contributes to the measured flux, so the geometric crossing time is scaled by
a fill factor

\begin{equation}
f = \min\!\left(\frac{A}{L\,w},\,1\right), \qquad
t_{\rm cross} = f\,\frac{L}{\dot\theta},
\end{equation}

where $L$ is the observed segment length (\texttt{Length}), $A$ the area of the
SAT-flagged spaxels (\texttt{Area}), and $w=12\arcsec$ the full width of the
masking strip (Section~\ref{sec:extractions}). The median fill factor is 0.27
(16th--84th percentile 0.18--0.41), set by the gaps between IFUs along a
streak's path; without it $t_{\rm cross}$ is overestimated and $g_{\rm inst}$
too faint by $\sim$1.4\,mag at the median. The instantaneous magnitude is then
\begin{equation}
g_{\rm inst} = g_{\rm AB} + 2.5\,\log_{10}\!\left(\frac{t_{\rm cross}}{360\,{\rm s}}\right),
\end{equation}
tabulated as \texttt{ginst}. Because $t_{\rm cross}\ll360$\,s, $g_{\rm inst}$ is
brighter than $g_{\rm AB}$: over the 475 identified streaks the correction has a
median of 6.2\,mag, ranging from 8.8\,mag for LEO to 5.3, 4.1 and 3.8\,mag for
MEO, HEO and GEO respectively, since $t_{\rm cross}$ scales inversely with
angular rate.

\subsection{Data Products}
\label{sec:catalog}

The final catalog, \texttt{HETDEX\_PDR1\_satellites.fits}, is distributed as a
single multi-extension FITS file containing 527 satellite streaks across 492
HETDEX PDR1 shots; five streaks initially classified as satellites whose spectra showed emission features
characteristic of meteors rather than reflected sunlight were removed prior to
analysis. Of the 527 streaks, 475 (90.1\%) are matched to a catalogued
object via the SGP4 identification pipeline described in
Section~\ref{sec:crossmatch}.

A binary-table extension lists one row per streak. Table~\ref{tab:catalog_columns} lists all columns and their descriptions. The first 17 columns record
the geometric, temporal, and photometric quantities measured from the data cubes
(Section~\ref{sec:extractions}): a flux-independent on-track centroid, the two
observed endpoints of the streak segment (which place the trail on the sky for
crossmatching), the projected segment length and position angle, SDSS-$g$
photometry, and the shot MJD and exposure time. Four further columns record the seeing (observation image quality) and the precise dither timing (Section~\ref{sec:crossmatch}). The remaining 22 columns
carry the orbital identification: the NORAD catalog number, object name, COSPAR
designator, object type, orbit class, match residuals, crossing epoch, slant
range, orbital elements, and instantaneous trail magnitude. Identification
columns are suppressed (masked) for the 52 unmatched streaks.

For the FITS file, three image extensions store, for every streak, the summed spectrum
(\texttt{SPECTRA}), its propagated $1\sigma$ uncertainty (\texttt{ERRORS}), and
the common wavelength grid (\texttt{WAVE}; 3470--5540\,\AA\ in 1036 channels of
2\,\AA), with row $i$ of each image extension corresponding to row $i$ of the
binary table. A fifth extension (\texttt{CANDIDATES}) lists the top five
identification candidates per streak for vetting ambiguous cases. Plain-text
(AAS machine-readable format, \texttt{HETDEX\_PDR1\_satellites.txt}) and CSV
(\texttt{HETDEX\_PDR1\_satellites.csv}) versions of the binary table are
provided alongside the FITS file. The catalog, analysis code, and per-shot
diagnostic plots---IFU footprints, masked-spaxel paths, and summed spectra---are
archived at \dataset[10.5281/zenodo.22962479]{https://doi.org/10.5281/zenodo.22962479};
the analysis pipeline is also maintained on GitHub\footnote{
\url{https://github.com/eriniocentric/hetdex_sats}}.

\begin{deluxetable}{ll}
\tablewidth{\columnwidth}
\tabletypesize{\scriptsize}
\tablecaption{Columns of the \texttt{HETDEX\_PDR1\_satellites.fits} binary-table
extension. \label{tab:catalog_columns}}
\tablehead{\colhead{Column} & \colhead{Description (unit)}}
\startdata
\multicolumn{2}{l}{\textit{Streak observation -- All 527 Rows Populated}} \\[2pt]
\texttt{ID}      & Streak identifier (0-based) \\
\texttt{Shot}    & HETDEX shot identifier (YYYYMMDDXXX) \\
\texttt{MJD}     & Shutter-open MJD of first dither, UTC (d) \\
\texttt{Texp}    & Exposure time per dither (s) \\
\texttt{RAcen}   & R.A.\ of streak centroid, J2000 (deg) \\
\texttt{DEcen}   & Decl.\ of streak centroid, J2000 (deg) \\
\texttt{RAsta}   & R.A.\ of first streak endpoint (deg) \\
\texttt{DEsta}   & Decl.\ of first streak endpoint (deg) \\
\texttt{RAend}   & R.A.\ of second streak endpoint (deg) \\
\texttt{DEend}   & Decl.\ of second streak endpoint (deg) \\
\texttt{Length}  & Observed segment length (arcsec) \\
\texttt{PA}      & Track position angle, E of N (deg) \\
\texttt{Area}    & Area from SAT-flagged spaxels (arcsec$^{2}$) \\
\texttt{Nifu}    & Number of IFUs contributing \\
\texttt{gmag}    & SDSS-$g$ AB magnitude (mag) \\
\texttt{gSB}     & SDSS-$g$ surface brightness (mag\,arcsec$^{-2}$) \\
\texttt{SNR}     & Mean S/N\,pix$^{-1}$, 3800--5500\,\AA \\[4pt]
\texttt{FWHM}    & Seeing FWHM from VIRUS guider (arcsec) \\
\texttt{Dither}  & Dither number (1--3) in which streak was detected \\
\texttt{MJDopen} & Shutter-open MJD for that dither, UTC (d) \\
\texttt{MJDclos} & Shutter-close MJD for that dither, UTC (d) \\[4pt]
\multicolumn{2}{l}{\textit{Orbital Identification -- Blank for 52 Unmatched Streaks}} \\[2pt]
\texttt{NORAD}   & NORAD catalog number \\
\texttt{Name}    & Object name \\
\texttt{COSPAR}  & COSPAR international designator \\
\texttt{Type}    & Object type (PAYLOAD, ROCKET BODY, DEBRIS) \\
\texttt{Owner}   & Owning nation or operator code \\
\texttt{Launch}  & Launch date (YYYY-MM-DD) \\
\texttt{Constel} & Constellation membership \\
\texttt{Class}   & Orbit class (LEO, MEO, GEO, HEO) \\
\texttt{Resid}   & Cross-track offset from propagated arc (arcsec) \\
\texttt{dPA}     & Position-angle residual (deg) \\
\texttt{Uniq}    & 1 if best match is unambiguous \\
\texttt{MJDx}    & MJD of closest approach, UTC (d) \\
\texttt{TLEage}  & Shot epoch minus TLE epoch (h) \\
\texttt{Range}   & Topocentric slant range (km) \\
\texttt{Height}  & Height above Earth ellipsoid (km) \\
\texttt{Rate}    & Topocentric angular rate (arcsec\,s$^{-1}$) \\
\texttt{Incl}    & Orbital inclination (deg) \\
\texttt{Perigee} & Perigee height (km) \\
\texttt{Apogee}  & Apogee height (km) \\
\texttt{Period}  & Orbital period (min) \\
\texttt{ginst}   & Instantaneous SDSS-$g$, trail-rate and fill corrected$^{a}$ (mag) \\
\texttt{IDsrc}   & ID source (\texttt{spacetrack} or \texttt{satchecker}) \\
\enddata
\tablenotetext{a}{\texttt{ginst} normalised to 360\,s reference exposure and
corrected for the IFU fill factor; see Section~\ref{sec:photometry}.}
\tablecomments{Spectral flux (\texttt{SPECTRA}, \texttt{ERRORS} HDUs) in
$10^{-17}\,\mathrm{erg\,s^{-1}\,cm^{-2}\,\AA^{-1}}$. \texttt{Dither}, \texttt{MJDopen} and \texttt{MJDclos} are blank for the 6
streaks that could not be assigned to a single dither in the track list. The \texttt{CANDIDATES} HDU lists the five best-scoring identifications per
streak, for vetting the 4 ambiguous matches.
Also available as \texttt{HETDEX\_PDR1\_satellites.txt} (AAS MRT format). }
\end{deluxetable}

\section{Results}
\label{sec:results}

\begin{figure*}[t]
    \centering
    \includegraphics[width=0.97\textwidth]{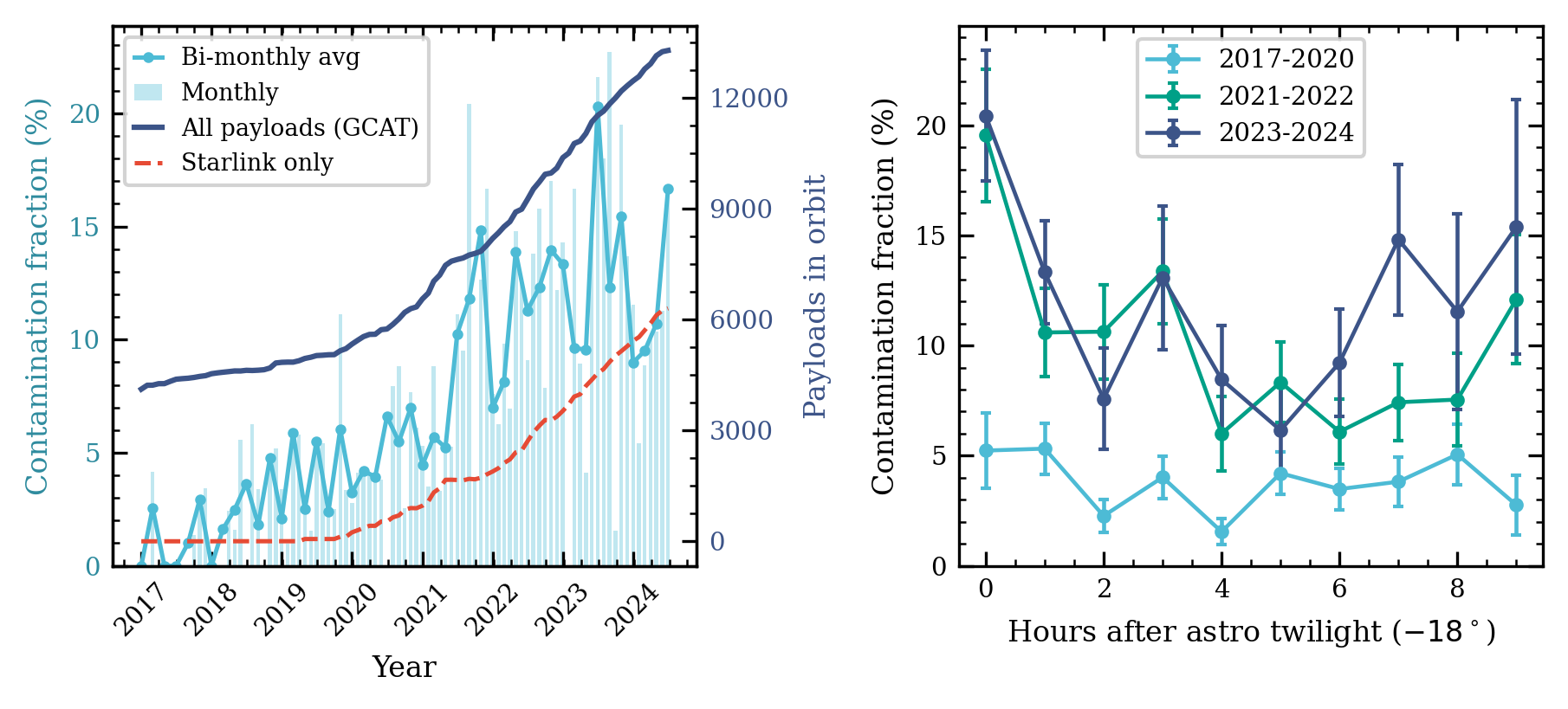}
    \caption{Satellite contamination in HETDEX PDR1. Left: growth over the survey. Light blue bars show the monthly fraction of observations containing at least one satellite streak and light blue points the bi-monthly average (left axis); the dark blue and red curves give the number of all active payloads and Starlink payloads in orbit, respectively, from GCAT \citep{GCAT} (right axis). The contamination fraction rises from a few percent early in the survey to monthly values above 20\% by 2023, broadly tracking the growth of the low-Earth-orbit population. Right: contamination fraction vs. hours after astronomical twilight (when the Sun falls 18 deg below the horizon for HET), split into survey epochs (2017--2020, 2021--2022, 2023--2024; one-hour bins, binomial errors). Latter epochs show the highest contamination values near twilight, but the values remain significant all night.}
    \label{fig:contam}
\end{figure*}

We identify 527 satellite streak records across 492 unique HETDEX observations, out of 6778 total observations in PDR1, corresponding to an overall contamination fraction of 7.3\%. The observations span MJD 57834--60522. The contamination fraction evolves significantly over this period, reaching a peak monthly value of 22.7\% in July 2023.

While satellite streaks affect a substantial fraction of observations, the sky area they contaminate is small. Summing the masked spaxels over all 527 streaks yields a total flagged area of $1.03\times10^{6}$ arcsec$^2$ (0.08 deg$^2$), corresponding to just 0.09\% of the 86.67 deg$^2$ PDR1 effective survey area. Individual streaks mask a mean area of $\sim$2000 arcsec$^2$, roughly 1\% of the IFU-covered area in a typical observation. Because flagged spaxels may be flagged for other reasons as outlined in \citet{hetdex_pdr1}, this represents an upper limit on the survey area lost to satellite contamination. 

As noted in Section\,\ref{sec:extractions}, the masked area is not an intrinsic property of the satellite but is set by the adopted masking half-width, which defines how far perpendicular to the fitted track fibers are flagged (a default of $6\arcsec$; \texttt{streaksize}). Because the flagged region is a strip of width twice this masking half-width, the total masked area scales approximately linearly with it, and the values quoted here should be interpreted with respect to the $6\arcsec$ masking adopted by the HETDEX pipeline and not a true measure of affected area.

\subsection{Growth of Satellite Contamination}
\label{sec:contamination}

Figure~\ref{fig:contam} summarizes the rate of contamination throughout the PDR1 survey period. The left panel shows the monthly and bi-monthly contamination fraction rising from a few percent in 2017 to monthly values exceeding 20\% by 2023, broadly tracking the growth of payloads in orbit from the General Catalog of Artificial Space Objects (GCAT\footnote{\url{https://planet4589.org/space/gcat/web/cat/index.html}}; \citealt{GCAT}). However, the fractional contamination does not rise as steeply as the total number of satellites in orbit, suggesting that geometric and observational factors limit the probability that any given satellite intersects a HETDEX field of view during an exposure. As predicted by modeling studies by \citet{Hainaut2020, Tyson2020}, our contamination fraction plateaus around 20--25\%.

The right panel shows the contamination fraction as a function of hours after astronomical twilight, split into three survey epochs. In the most recent epoch (2023--2024) the fraction peaks at $20.4\%$ in the first hour after twilight and averages $10.5\%$ over the remainder of the night---roughly a factor of 2 enhancement---dipping to a midnight minimum before rising again toward morning twilight. This twilight excess has strengthened over the survey: the 2021--2022 epoch shows a comparable first-hour peak ($19.5\%$) but a lower nighttime baseline ($8.9\%$), while in 2017--2020 the fraction is nearly flat, with only a weak enhancement in the first hour ($5.2\%$ versus a $3.5\%$ baseline). The growth of the post-twilight peak thus tracks the increasing LEO satellite population, even as streaks remain detectable at all hours of the night.

\subsection{Photometric Properties}
\label{sec:photometric}
\begin{figure*}[t]
    \centering
    \includegraphics[width=0.95\textwidth]{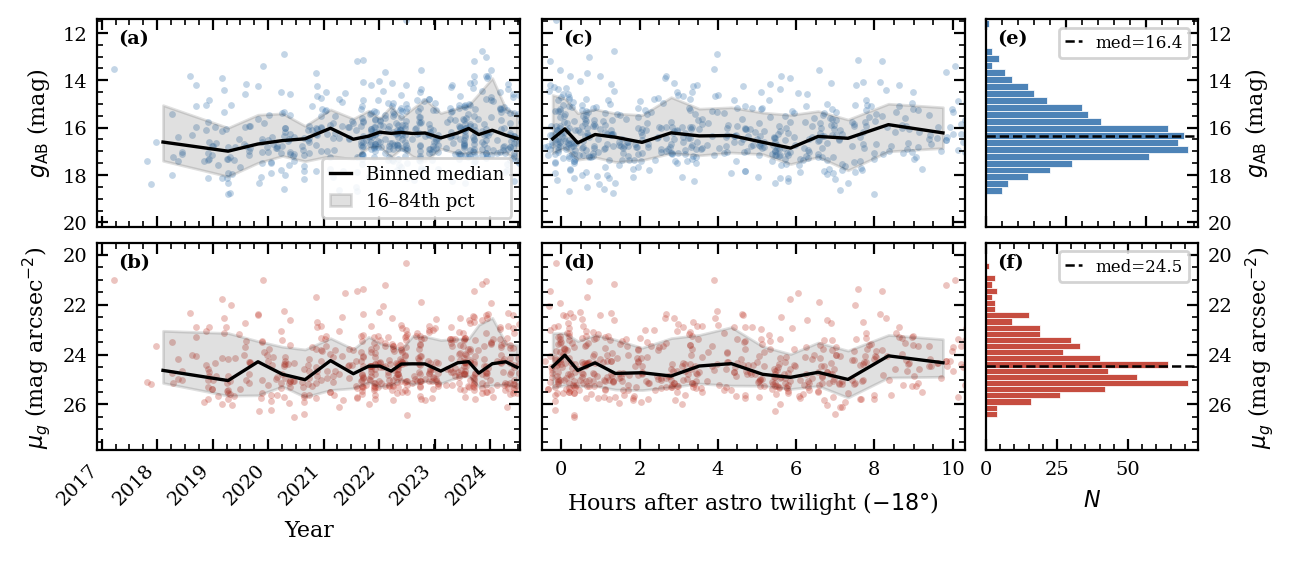}
    \caption{
    Photometric properties of satellite streaks measured from HETDEX spectra.
    Panels (a) and (b) show SDSS $g$-band magnitude and surface-brightness as a function of year.
    Panels (c) and (d) show the same quantities as a function of time after astronomical twilight.
    Points show individual satellite streaks, black curves show binned medians, and shaded regions indicate the
    16th--84th percentile range.
    Panels (e) and (f) show the corresponding distributions, with dashed lines marking the median values
    of $g_{\rm AB}=16.4$ and $\mu_g=24.5$ mag arcsec$^{-2}$.
    The broad distributions reflect variation in satellite altitude, phase angle, apparent trail length, and observing geometry.
    }
    \label{fig:gmag_trends}
\end{figure*}

For each satellite streak, we measure an integrated SDSS $g$-band magnitude and a mean surface brightness as described in Section\,\ref{sec:extractions}. The brightness distribution has a median value of $g_{\rm AB}=16.4$ with a standard deviation of 1.1 mag, while the surface brightness distribution has a median value of $\mu_g=24.5$ mag arcsec$^{-2}$ with a standard deviation of 1.0 mag arcsec$^{-2}$.

These values include the dither-dilution correction of
Section\,\ref{sec:extractions}, but they remain exposure-averaged: a
low-Earth-orbit satellite crosses the $22\arcmin$ field in $\lesssim1$\,s and
dwells on any individual spaxel for only $\sim$0.2\,ms, whereas the flux
calibration normalizes the collected photons by the full $\sim$360\,s exposure.
The magnitudes tabulated here therefore measure the light a satellite
contributes to a given exposure---the physically relevant quantity for
assessing survey impact---rather than its instantaneous brightness, which we
return to in Section\,\ref{sec:g_inst}.

Figure~\ref{fig:gmag_trends} shows the measured magnitudes and surface brightnesses as a function of both year and time after astronomical twilight defined as the time when the Sun falls 18 deg below the horizon for the HET. The distributions are broad, reflecting variations in satellite altitude, phase angle, intrinsic reflectivity, and observing geometry, but show no strong monotonic evolution over the HETDEX observing baseline. These measurements indicate that satellite streaks are typically bright in integrated flux but have much lower surface brightness when distributed across the HETDEX field of view.

We do not observe a strong concentration of brighter streaks near twilight but there is an increase in frequency (as discussed in Section\,\ref{sec:contamination}). While satellite visibility is often assumed to peak at small Sun--target angles, the HETDEX sample shows a broad distribution in both measured magnitude and surface brightness over several hours after astronomical twilight. This behavior is enabled in part by the large collecting area of the HET, which allows for detection of satellite streaks over a wider range of brightness and observing geometries than would be accessible to smaller-aperture facilities.

\subsection{Spectral Properties}
\label{sec:spectral}

\begin{figure*}[t]
    \centering
    \includegraphics[width=0.97\textwidth]{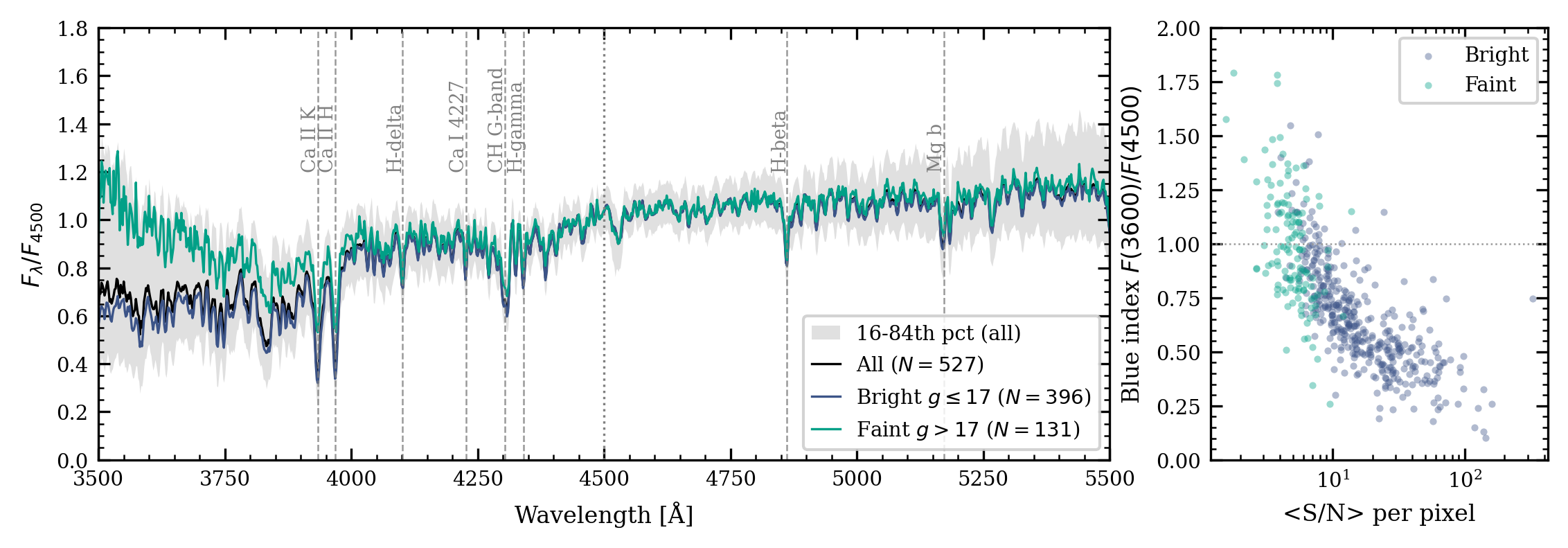}
    \caption{Spectral properties of the HETDEX satellite-streak sample. Left: median stacked spectra, normalized to unity at 4500\,\AA, for the full sample (black), the bright subsample ($g\le17$; dark blue), and the faint subsample ($g>17$; teal). Dashed vertical lines mark the wavelengths of prominent solar Fraunhofer features (Ca,II H\&K, the CH G-band, the Balmer series, and Mg\,b) for reference; the smooth continua are consistent with reflected sunlight. Right: the per-streak blue index, the ratio of flux near 3600\,\AA\ to that near 4500\,\AA, versus mean spectral signal-to-noise for the bright (dark blue) and faint (teal) sub-samples.
    }
    \label{fig:satellite_spectra}
\end{figure*}

In the left panel of Figure~\ref{fig:satellite_spectra}, we display median spectra for the full sample in black as well as for sub-samples split by brightness, defined as bright ($g\le17$ in dark blue) and faint ($g>17$ in teal). All spectra are normalized at 4500\,\AA\ to isolate differences in spectral shape independent of overall flux.  All stacks display clear solar spectrum features as marked in the figure indicating the spectra largely consist of reflected solar light. The faint subsample exhibits a steeper blue slope than the bright subsample shortward of 4500\,\AA. The right panel shows the excess versus signal-to-noise. The per-streak blue index $F(3600)/F(4500)$ increases toward low signal-to-noise. This change in colour for low S/N could be intrinsic if fainter sources have different reflectivity compared to brighter sources. But it could also be the result of summing many faint spaxels which are known to have a higher uncertainty in the blue regime. A complete analysis would incorporate properties of the specific satellites, mainly by identifying what metal is causing the solar reflection and the incidence angles. Another consideration is quantifying the sky-subtraction residuals (as is done in \citealt{davis2023b, weiss2024} for example).

\begin{figure*}
\centering
\includegraphics[width=\textwidth]{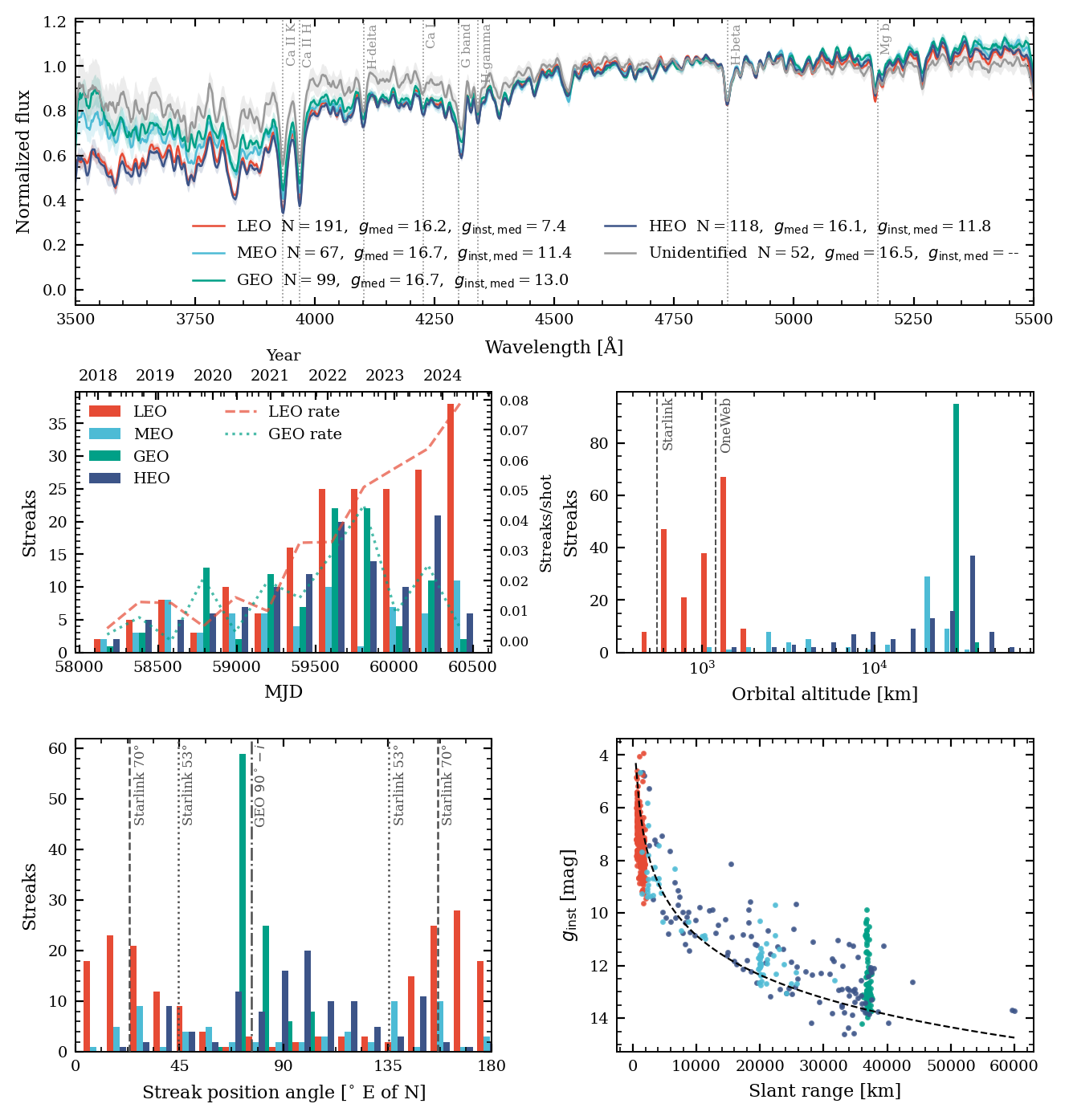}
\caption{%
Summary of the 475 identified satellite streaks in the HETDEX PDR1 catalog.
\textit{Top:} median-normalised streak spectra stacked by orbit class
(LEO, MEO, GEO, HEO, and the 52 unidentified streaks). Median measured $g$ and $g_{\rm inst}$ values are given for each identified class; the Unidentified class has no $g_{\rm inst}$ value.
\textit{Middle left:} identified streaks per semi-annual bin vs. epoch,
by orbit class. The contamination fraction (streaks per shot) for LEO and GEO 
populations is shown with red-dashed and green-dotted lines respectively.
\textit{Middle right:} height above the Earth ellipsoid at crossing, by
orbit class.
\textit{Bottom left:} streak position angle (folded to
$0\arcdeg$--$180\arcdeg$, east of north), by orbit class. Vertical lines
mark the ascending/descending azimuths of the Starlink 53\arcdeg\ (dotted)
and 70\arcdeg\ (dashed) shells at the HET latitude ($+30\fdg7$); the
dot-dashed line at $76\arcdeg$ marks $90\arcdeg-i$ for the median
inclination $i=14\arcdeg$ of the GEO-class identifications.
\textit{Bottom right:} instantaneous SDSS-$g$ magnitude vs.\ topocentric
slant range, by orbit class. The dashed curve is the inverse-square locus
anchored to the sample median $g(550\,\mathrm{km})$.
\label{fig:crossmatch_summary_panel}%
}
\end{figure*}

\subsection{Orbital Identification}
\label{sec:identification}

Of the 527 satellite streaks in the HETDEX PDR1 catalog, 475 (90.1\%) are
matched to a catalogued object: 468 via SGP4 propagation of Space-Track
archival TLEs and 7 via the IAU CPS SatChecker archive for streaks that the
SGP4 pipeline searched and did not resolve. Identifications therefore rest
almost entirely (98.5\% of matches) on a single, uniformly applied element
source; the 7 SatChecker-derived rows are flagged by \texttt{IDsrc} and were
accepted only after passing the same geometric tolerances as every other
match. The 52 unmatched streaks are retained in the catalog with
identification columns empty.

For the 475 matched streaks, the median cross-track offset between the
observed streak midpoint and the best-fitting propagated great-circle arc is
13.9\arcsec\ (90th percentile 66.1\arcsec), well within the 900\arcsec\
tolerance adopted to account for TLE along-track uncertainty. The median
position-angle residual is $0.08\arcdeg$ (90th percentile $0.54\arcdeg$), and
471 of 475 matches (99.2\%) are unambiguous. As a purity check, five matched
streaks are assigned orbits whose SGP4 propagation places the object in
Earth's umbra at crossing time, giving a false-positive floor of
$\lesssim$1.1\%. 

As an independent check, a random subset of 180 of the 468 Space-Track matches
was reidentified through the IAU CPS SatChecker API \citep{satchecker}, which
propagates its own element archive. Of the 166 SatChecker resolved --- 14
predate its archive --- 161 agree (97.0\%), with comparable residuals (median perpendicular offset
12.6\arcsec\ against our 13.9\arcsec). Of the five disagreements, three have
the Space-Track candidate within 15\arcsec\ of the track against
183--878\arcsec\ for SatChecker's; the other two carry cross-track residuals
above 300\arcsec, the top 1.3\% of our distribution, and are the only plausible
misidentifications.

A full set of
diagnostic figures for each crossmatch (\texttt{crossmatch\_diagnostic\_figures.zip})
is archived on Zenodo (doi:\dataset[10.5281/zenodo.22962479]{https://doi.org/10.5281/zenodo.22962479}).

\subsubsection{Orbital Properties}

By object type, the orbital identifications consist of 254 payloads (53.5\%), 172 rocket
bodies (36.2\%), and 49 debris objects (10.3\%). By orbit class, 191 matches
are LEO (40.2\%), 67 MEO (14.1\%), 99 GEO (20.8\%), and 118 HEO (24.8\%). The
sensitivity provided by HET's large aperture and HETDEX's 360\,s exposure time 
result in similar ranges of observed magnitudes for GEO and LEO populations even 
though the lower-altitude LEOs are visually brighter by eye.

By constellation, 34 matches (7.2\%) are
Starlink, 50 (10.5\%) Cosmos, 21 (4.4\%) OneWeb, 15 (3.2\%) Globalstar, and
340 (71.6\%) are unaffiliated or uncatalogued objects. The Cosmos entry collects
objects carrying the Soviet and Russian Kosmos cover designation, a series
spanning launches from 1965 to 2019 and encompassing both defunct military
payloads and operational GLONASS navigation satellites; it is a designation
rather than a constellation brand in the sense of Starlink or OneWeb. That such
objects outnumber Starlink among the identifications reflects the 2017--2024
baseline of PDR1: the survey predates the constellation by two years, and no
Starlink streak appears before 2021.

In the top panel of Figure\,\ref{fig:crossmatch_summary_panel}, median normalized 
spectra for each orbital class are provided, GEO and MEO types are fainter on average 
and are slightly bluer, but this may be primarily due to lower S/N as discussed in Section\,\ref{sec:spectral}. 

A breakdown of orbit class as a function of time is provided in the middle-left panel of 
Figure\,\ref{fig:crossmatch_summary_panel}. There is a clear rise in LEO objects (red)
over time. GEO objects show an increase in counts in 2022-2023 but when adjusted for observation
frequency, the rate of GEO objects is relatively flat over time (dotted green line) while
the rate of LEO object contamination increases markedly (red dashed line).

The launch dates of the identified objects span 1963--2024 with a median of
1999, so the typical streak is produced by hardware $\sim$23\,yr old at the
time of observation (16th--84th percentile 3--39\,yr). Just over half (53\%)
were launched before 2000 and two-thirds before 2010; only 69 objects (14.5\%)
date from the 2020s. The distribution is two-peaked: a broad maximum through
the 1980s and 1990s (190 objects, 40\%), dominated by spent upper stages and
defunct payloads, and a second rise after 2015 (119 objects, 25\%) driven by
the LEO constellations. Launch epoch also varies with object type: Rocket bodies
have a median launch year of 1994, against 1999 for payloads and 2008 for
debris.  It varies with orbit class as well, with the geostationary objects the oldest at
observation (median age 30\,yr). The payloads are the most broadly
distributed population (16th--84th percentile 1983--2021), combining decades of
abandoned geostationary satellites with recently launched constellation
members. What HETDEX records is therefore the accumulated inventory of six
decades of launch activity rather than a snapshot of the current operational
fleet.

The dominance of older, spent objects in our sample is not
evidence that modern satellite constellations pose a lesser threat to
astronomy. We identify two selection effects unique to HETDEX. First,
HETDEX's PDR1 baseline spans 2017--2024, giving pre-constellation-era
debris and rocket bodies far more cumulative opportunities to cross the
field than recently launched satellites, which make up only 35\% of
identifications. Second, HETDEX's sensitivity and $\sim$6\,min exposures
result in similar exposure-averaged magnitudes for LEO and GEO objects
(median $g_{\rm AB}=16.2$ versus 16.7), even though their instantaneous
brightness while crossing the field varies widely 
($g_{\rm inst}=7.4$ versus 13.0, a 5.6\,mag difference). If we consider the
orbit classes separately, as in the middle-left panel of
Figure~\ref{fig:crossmatch_summary_panel}, it is clear the brighter LEO
population is increasing over time. This matches observations from imaging
surveys with less sensitive instruments, such as 32 million photometric
observations from the Mini-MegaTORTORA network \citep{Mallama2026}, and
what naked-eye observers see: The number of satellites visible in the sky
is increasing greatly over time and continues to contaminate observations
at greater levels.

The altitude distribution (middle-right panel of
Figure\,\ref{fig:crossmatch_summary_panel}) shows that LEO matches span $406$--$1719$\,km and are double-peaked:
43 streaks form a spike at 500--600\,km, 33 of them Starlink
(median 542\,km), and 66 lie between 900 and 1300\,km, including all 21 OneWeb
matches (median 1203\,km). Only 46 streaks (9.7\%) fall in the inner Van Allen
belt at 2000--10{,}000\,km, which operators avoid. The MEO population peaks at
the GLONASS shell (median altitude $19{,}102$\,km, median period $674.6$\,min). The geostationary belt is the sharpest feature in the panel with 99 GEO matches
at $35{,}915 \pm 267$\,km, $\sim$130\,km above the nominal geostationary radius,
consistent with a mixture of station-kept and drifting objects. The HEO class is
not a shell but launch hardware with a median eccentricity of $0.66$, a median perigee of $529$\,km, and a median apogee $\sim33{,}749$\,km, and 69 of 118 matches are rocket bodies.

Position angles (bottom-left) are folded to $0\arcdeg$--$180\arcdeg$ east of
north, since a streak records an axis rather than a direction of travel. For
inclination $i$ observed from geodetic latitude $\phi$ the ground-track heading
satisfies $\sin a = \cos i / \cos\phi$, giving position angles $a$ and
$180\arcdeg - a$; at $\phi = 30.681\arcdeg$, the Starlink $53\arcdeg$ and
$70\arcdeg$ shells predict $44.4\arcdeg/135.6\arcdeg$ and 
$23.4\arcdeg/156.6\arcdeg$ respectively (dotted and dashed lines). For the 312 matches with
$i > \phi$ the median offset from the nearer prediction is $14.4\arcdeg$, and
71\% agree to within $20\arcdeg$. LEO streaks avoid the east--west direction
(162 of 191 at $\mathrm{PA} < 45\arcdeg$ or $> 135\arcdeg$; only 7 between
$60\arcdeg$ and $100\arcdeg$), whereas GEO streaks are the most concentrated
population in the catalog: 84 of 99 lie between $70\arcdeg$ and $90\arcdeg$ with
a median of $75.3\arcdeg$, matching the $90\arcdeg - i = 76.0\arcdeg$ expected
for the median GEO inclination of $14.0\arcdeg$ (dash-dotted line). MEO position
angles are distributed uniformly.

\subsubsection{Instantaneous Magnitudes}
\label{sec:g_inst}
The measured streak SDSS-$g$ magnitudes span $11.5$--$18.8$\,mag (median
16.4\,mag) and observed lengths range from 0.6\arcmin\ to 17.2\arcmin\ (median
11.1\arcmin). As trail-integrated values they compress the orbit classes
together, with class medians differing by only $0.52$\,mag from LEO ($16.22$) to
GEO ($16.74$) despite a factor of $\sim$35 in distance: the transit-time
correction of Section\,\ref{sec:photometry} very nearly cancels the
inverse-square penalty, which rises over the same interval from $1.6$ to
$9.1$\,mag. This is why HETDEX records high-orbit debris at all. 

When we correct for transit time, the instantaneous magnitudes span $g_{\rm inst} = 3.9$--$14.6$\,mag (median
$10.0$\,mag) and separate the classes cleanly, with medians of $7.41$ (LEO),
$11.36$ (MEO), $11.81$ (HEO) and $12.98$ (GEO) --- a range of $5.6$\,mag
against $0.67$\,mag for the measured values. Our GEO detections, at
$g_{\rm inst}\sim13$, are well below the naked-eye visibility brightness limit targeted by the IAU \citep{IAU2024}, indicating that even fully
compliant satellites can register as contamination in deep, large-aperture-telescope
surveys.
 
The bottom-right panel of Figure\,\ref{fig:crossmatch_summary_panel} plots
$g_{\rm inst}$ against slant range, with the inverse-square locus
$5\log_{10}(d/550\,\mathrm{km})$ overlaid. HET's fixed-altitude design confines
the detections to a narrow band of elevation ($48.6\arcdeg$--$61.5\arcdeg$,
median $54.8\arcdeg$), so air mass and topocentric geometry are nearly constant
across the sample, and slant range tracks orbital altitude almost
deterministically. The trail-integrated magnitudes carry almost no distance
information (Spearman $\rho = 0.18$ against $\log d$); the point-source
magnitudes follow the locus with $1.4$\,mag rms.
 
Normalising to 550\,km gives a quantity comparable to published photometry. The
matched sample has $g(550\,\mathrm{km}) = 4.55$\,mag at the median (16th--84th
percentile: $3.25$--$6.12$), rising from $3.86$ for GEO to $5.91$ for LEO: the
high-orbit objects are intrinsically the brightest, as expected for
geostationary buses and spent upper stages. The 34 Starlink matches give
$5.95$\,mag ($5.23$--$6.94$), in agreement with the mean
$V(550\,\mathrm{km}) = 5.92 \pm 0.04$ reported for VisorSat-design spacecraft by
\citet{Mallama2021}. That agreement is an end-to-end check on the photometric
chain, since the published value comes from visual photometry of resolved
passes rather than from trail spectroscopy. The residual scatter about the
locus is driven by solar phase angle, which is uncontrolled here, and by
tumbling. The geostationary objects are the cleanest case: their slant ranges
span only $0.10$\,mag of distance, so their $4.3$\,mag vertical extent in the
panel is intrinsic almost by construction.

\subsubsection{Unmatched Streaks}

The 52 unmatched streaks are not attributable to gaps in the element-set
archive: every observing night in PDR1 is covered by the TLE cache, and each
unmatched streak was propagated against a median of $\sim$21{,}000 catalogued
objects, of which $\sim$92 passed the coarse positional filter --- statistically
indistinguishable from the matched sample. They are also distributed across the full survey baseline rather than
concentrated in any epoch, with unmatched fractions ranging from 3.1--19\% with
no trend, and their median $g$ of 16.5\,mag sits between those of the LEO and
GEO classes. Their stacked spectrum
(top panel Figure~\ref{fig:crossmatch_summary_panel}) shows the same solar Fraunhofer
absorption features, at comparable depth, as the stacks of the four identified
orbit classes, suggesting the unmatched streaks are reflected sunlight from real objects,
not instrumental artifacts or cosmic rays. We note that there appears a modest flux excess compared to the matched sample blueward of 4200\,\AA\, which could point to a different reflective material.

\section{Conclusions}

We present a spectroscopic analysis of satellite streaks in HETDEX PDR1,
identifying 527 streaks across 492 observations. The streaks have a median
$g_{\rm AB}\sim16.4$ and $\mu_g\sim24.5$\,mag\,arcsec$^{-2}$, but because
each $\sim$360\,s exposure far exceeds the satellite transit time, these are
strict lower limits: Instantaneous magnitudes are brighter by 6\,mag on
average, with class medians of $g_{\rm inst}=7.4$ (LEO), $11.4$ (MEO),
$11.8$ (HEO), and $13.0$ (GEO). The spectra are consistent with reflected
sunlight. Despite affecting many observations, the streaks mask only
$\sim$0.09\% of the survey area and can be sufficiently
excluded from survey data.

The fraction of affected observations rises from a few percent early in the survey to monthly values above 20\% by 2023. Contamination is most frequent near twilight but remains significant at all hours, and this enhancement has strengthened over the survey---being nearly flat in 2017--2020, reaching $\sim$20\% in the first hour after twilight by 2023--2024. The per-streak brightness shows no significant evolution, indicating that the rising impact is driven by the growing frequency of contamination rather than by any change in the streaks themselves.

Propagating archival TLEs identifies the object responsible
for 475 of the streaks (90.1\%). Only 35\% were launched since 2010; the
rest are spent rocket bodies, catalogued debris, and spacecraft launched
decades earlier, so the streaks trace six decades of accumulated launch
activity rather than any one constellation. This reflects HETDEX's baseline
observational coverage (from early 2017 to mid-2024) and sensitivity, not a
reprieve from constellation growth: The brighter LEO population is
demonstrably increasing over time. Nearly half (46\%) are in geostationary
or highly elliptical orbits---a selection effect of the telescope rather
than the sky, since the Hobby-Eberly Telescope's large 10\,m aperture and
HETDEX's moderately long, $\sim$6\,min exposures bring distant, slow-moving
GEO and HEO objects within reach at nearly the same observed brightness as
much closer LEO objects.


\begin{acknowledgments}

HETDEX is led by the University of Texas at Austin McDonald Observatory and Department of Astronomy with participation from the Ludwig-Maximilians-Universit\"at M\"unchen, Max-Planck-Institut f\"ur Extraterrestrische Physik (MPE), Leibniz-Institut f\"ur Astrophysik Potsdam (AIP), Texas A\&M University, The Pennsylvania State University, Institut f\"ur Astrophysik G\"ottingen, The University of Oxford, Max-Planck-Institut f\"ur Astrophysik (MPA), The University of Tokyo, and Missouri University of Science and Technology.

Observations for HETDEX were obtained with the Hobby-Eberly Telescope (HET), which is a joint project of the University of Texas at Austin, the Pennsylvania State University, Ludwig-Maximilians-Universit\"at M\"unchen, and Georg-August-Universit\"at G\"ottingen. The HET is named in honor of its principal benefactors, William P. Hobby and Robert E. Eberly. 

The Visible Integral-field Replicable Unit Spectrograph (VIRUS) was used for HETDEX observations. VIRUS is a joint project of the University of Texas at Austin, Leibniz-Institut f\"ur Astrophysik Potsdam (AIP), Texas A\&M University (TAMU), Max-Planck-Institut f\"ur Extraterrestrische Physik (MPE), Ludwig-Maximilians-Universit\"at München, Pennsylvania State University, Institut f\"ur Astrophysik G\"ottingen, University of Oxford, and the Max-Planck-Institut f\"ur Astrophysik (MPA). In addition to Institutional support, VIRUS was partially funded by the National Science Foundation, the State of Texas, and generous support from private individuals and foundations.

The authors acknowledge the Texas Advanced Computing Center (TACC) at The University of Texas at Austin for providing high performance computing, visualization, and storage resources that have contributed to the research results reported within this paper. URL: http://www.tacc.utexas.edu

We acknowledge RAIC Labs (\url{https://raiclabs.com}) for providing access to the RAIC platform, which enabled the efficient classification and identification of satellite streaks within the HETDEX dataset.

This research made use of data and/or services provided by the International Astronomical Union Centre for the Protection of the Dark and Quiet Sky (IAU CPS) SatHub. The development of SatChecker has been supported by the National Science Foundation (NSF) under grant numbers: AST 2332735 and AST 2332736. SatChecker is hosted at the NSF NOIRLab. Any opinions, findings, and conclusions or recommendations expressed in this material are those of the author(s) and do not necessarily reflect the views of the NSF, IAU, NSF NOIRLab, SKAO, ESO, or any host or member institution of the IAU CPS.

\textbf{AI Disclosure:} This work made use of Claude (Sonnet~4.5, Opus~4.5/4.8/5; \citealt{claude}), a large language model (LLM), for literature search, text editing, and developing and documenting parts of the analysis code described in Sections~\ref{sec:extractions} and Section~\ref{sec:crossmatch}. Photometric extraction (Section~\ref{sec:extractions}) used LLM context grounded in the HETDEX PDR1 paper \citep{hetdex_pdr1} and its associated \texttt{dexcube} repository (\url{https://github.com/HETDEX/dexcube}). The matching pipeline (Section~\ref{sec:crossmatch}), built on \texttt{sgp4} propagation of archival elements, was validated by cross-checking a subset of matches against \texttt{satchecker}, an independent identification service drawing on a separate element archive; diagnostic figures for each fit, available at \url{https://github.com/eriniocentric/hetdex_sats} and archived at \dataset[10.5281/zenodo.22962479]{https://doi.org/10.5281/zenodo.22962479}, provide additional validation of the crossmatching method. All code, results, and text were reviewed by the authors, who take full responsibility for the content and accuracy of this paper.

\end{acknowledgments}

\facilities{HET}

\software{astropy \citep{astropy:2013, astropy:2018},
          RAIC Labs \citep{raiclabs},
          hetdex-api \citep{hetdex_api},
          dexcube \citep{hetdex_pdr1},
          speclite \citep{speclite_pkg},
          astroplan \citep{astroplan},
          sgp4 \citep{sgp4pkg},
          satchecker \citep{satchecker},
          Claude \citep{claude},
          }

\bibliography{hetdex}{}
\bibliographystyle{aasjournalv7}

\end{document}